\documentclass[preprint,nopacs,preprintnumbers,amsmath,amssymb]{revtex4}
\usepackage{graphicx}

\usepackage{amsmath}
\usepackage{amssymb}
\usepackage{bm}
\usepackage{float}

\begin{document}
\title{Exact and LDA entanglement of tailored densities in an interacting one-dimensional electron system}

\author{J P Coe$^{1,2}$} 
\email{jpc503@york.ac.uk}

\author{I D'Amico$^1$}
\email{ida500@york.ac.uk}
\affiliation{
$^1$ Department of Physics, University of York, York YO10 5DD, United Kingdom.\\
$^2$ Department of Mathematics, University of York, York YO10 5DD, United Kingdom.}

\begin{abstract}
We calculate the `exact' potential corresponding to a one-dimensional interacting system of two electrons with a specific, tailored density.   We use one-dimensional density-functional theory with a local-density approximation (LDA) on the same system and calculate densities and energies, which are compared with the `exact' ones.  The `interacting-LDA system'\cite{COE} corresponding to the LDA density is then found and its potential compared with the original one. Finally we calculate and compare the spatial entanglement of the electronic systems corresponding to the interacting-LDA and original interacting system.    
\end{abstract}
\maketitle
\section{Introduction}

The accuracy of the entanglement calculated using approximations is  important when modelling the suitability of electron systems for use as quantum information devices.  Investigating this accuracy can also reveal more information about an approximation and perhaps ways of improving it. We look at the ground state (GS) of a one-dimensional (1D) contact-interacting system of two electrons. The electronic spins are maximally entangled (singlet state), so we focus on the spatial entanglement\cite{COE}. We model the system approximately using  density-functional theory (DFT)\cite{HK} with the one-dimensional analogue of the local-density approximation (LDA)\cite{MAGYAR04}.  
 DFT gives the GS density, not the many-body wave-function, and although in theory all GS properties are expressible as functionals of the GS density, it is not known how to calculate the entanglement this way. An alternative route is to find the (contact) interacting LDA system (i-LDA)\cite{COE} that gives the LDA density. We may now access the entanglement relatively easily from the i-LDA GS  many-body wavefunction\cite{COE}. By the Hohenberg-Kohn theorem\cite{HK} the GS density uniquely determines the external potential for a given form of interaction, then the i-LDA system must be the unique contact-interacting system that corresponds exactly to the LDA density.   In \cite{Coe2} we introduced the iterative scheme 

\begin{eqnarray}
v^{i+1}_{\text{ext}}(\bm{r_{1}})=\frac{1}{n_i(\bm{r_{1}})}|E_{i}|[n_{i}(\bm{r_{1}})-n^{\text{target}}(\bm{r_{1}})]
+v^{i}_{\text{ext}}(\bm{r_{1}})
\label{eq:bothscheme}
\end{eqnarray}

to find the potential $v_{\text{ext}}$ of the interacting system that reproduced a density arising from a single particle equation.  This scheme may also be used in a more general way, i.e. to find the interacting system that gives an experimental density, or to design a potential to give a desired (target) density. The latter application could be important in the future for the design of nano-devices.  Hence in this contribution we illustrate the usefulness of the iterative scheme (Eq.~\ref{eq:bothscheme}) in finding the numerically `exact' potential that gives a target density in a simple one dimensional two electron system of Hamiltonian $H=\delta(x_{1}-x_{2})+\sum_{i=1,2}\left[-\frac{1}{2}\frac{d^{2}}{dx_{i}^{2}}+v_{\text{ext}}(x_{i})\right]$ (atomic units).
We also check the scheme by showing that it can correctly reproduce a pre-chosen potential from  the corresponding density.  We then apply DFT and the LDA using the `exact' potential and again use the scheme to show how close the i-LDA potential is to the exact and to calculate the approximation to the entanglement corresponding to using the LDA.

To apply the iterative scheme Eq.~\ref{eq:bothscheme} 
to the tailored GS density and find the potential for the system,  we expand the interacting GS using the single particle eigenvectors corresponding to $v^{i}_{\text{ext}}$. This basis is changed at every step and the components of the GS are symmetrised.  This means that the calculations become much longer but  tractable for a one dimensional system with a contact interaction.  By changing the basis at each step we can access a much larger set of wave-functions with a small basis size so in theory allowing us to approximately reproduce any v-representable density. For fast calculations we use a single particle basis of $10$ thereby giving a symmetric two particle basis of $55$.

We investigate densities on the range $[-4:4]$ with $200$ mesh points and start with the trial potential $v^{1}_{\text{ext}}(x)=0.0001x^{2}$. This produces a relatively spread out density on this range so reduces the risk of the iterative scheme exploding at the start due to the denominator being too close to zero.
The error is quantified using $\frac{1}{N}\sum_{i=1}^{N}|n^{target}(x_{i})-n^{trial}(x_{i})| $ where $N$ is the number of mesh points ($200$ in this case).  The potential that gives the density to high enough accuracy is then used in the $1D$ DFT scheme to produce the LDA GS density.
\section{One-dimensional DFT using LDA}
We use the one-dimensional DFT for contact interacting fermions of Magyar and Burke \cite{MAGYAR04} with their local-density approximation.  DFT seeks to express the total GS energy as a functional of the density $n(r)$,
$E[n]=T_{NI}[n]+U[n]+E_{xc}[n]+\int v_{ext}(x) n(x)dx$.
This gives rise to the single-particle Kohn-Sham  equations
\begin{equation}
\left(-\frac{1}{2}\frac{d^{2}}{dx^{2}}+v_{ext}(x)+v_{H}(r;[n])+v_{xc}(r;[n]) \right )\phi_{i}(x)=\epsilon_i\phi_{i}(x),
\label{eq:KS}
\end{equation}
thereby allowing the density $n=\sum_{i}|\phi_{i}|^{2}$ to be efficiently computed.

Here $v_{H}=\frac{\delta U}{\delta n}$ and $v_{xc}=\frac{\delta E_{xc}}{\delta n}$.
 For a contact interaction $U=\int (n^{2}/2)dx$ so $v_{H}=n$. $E_{xc}$ can be written in the form $E_{xc}=\int e_{xc}(n) n(x) dx$
where $e_{xc}=e_{x}+e_{c}$. The results for a one dimensional homogeneous electron gas are used to give the local-density approximations   $e^{LDA}_{x}=-n/4$ and 
$e^{LDA}_{c}(n)=(an^{2}+bn)/(n^{2}+dn+e)$,
where $a=-1/24$, $b=-0.00436143$, $d=0.252758$, $e=0.0174457$ and we have used the  Pad\'{e} parametrization of Ref.~\cite{MAGYAR04}.

We then take the functional derivative of $E_{xc}$ to obtain $v_{xc}=v_{x}+v_{c}$. This gives $v_{x}(n)=-n/2$ and
\begin{equation}
v_{c}(n)=\left[1+n\left(\frac{2an+b}{an^{2}+bn}-\frac{2n+d}{n^{2}+dn+e}\right)\right]e_c(n).
\end{equation}

We may now use the LDA density with the iterative scheme Eq.~\ref{eq:bothscheme} to calculate the i-LDA potential. This is shifted by a constant to give equality of energy between the i-LDA and LDA system.  The procedure starting from a density with an unknown potential may be summarised as 
\begin{equation}n^{target} \xrightarrow{Eq.~\ref{eq:bothscheme}} v^{`exact'}_{ext} \xrightarrow{Eq.~\ref{eq:KS}} n^{LDA} \xrightarrow{Eq.~\ref{eq:bothscheme}} v^{i-LDA}_{ext}.\end{equation}

\section{System 1}

We first consider a known potential $v^{target}_{ext}(x)=0.02x+\sin(3x)+0.3x^{2}$ and the GS density $n^{target}$ arising from it. We wish to test the scheme's ability to reproduce the target density and potential.

\begin{figure}[!ht]
\begin{minipage}{18pc}
\includegraphics[width=18pc]{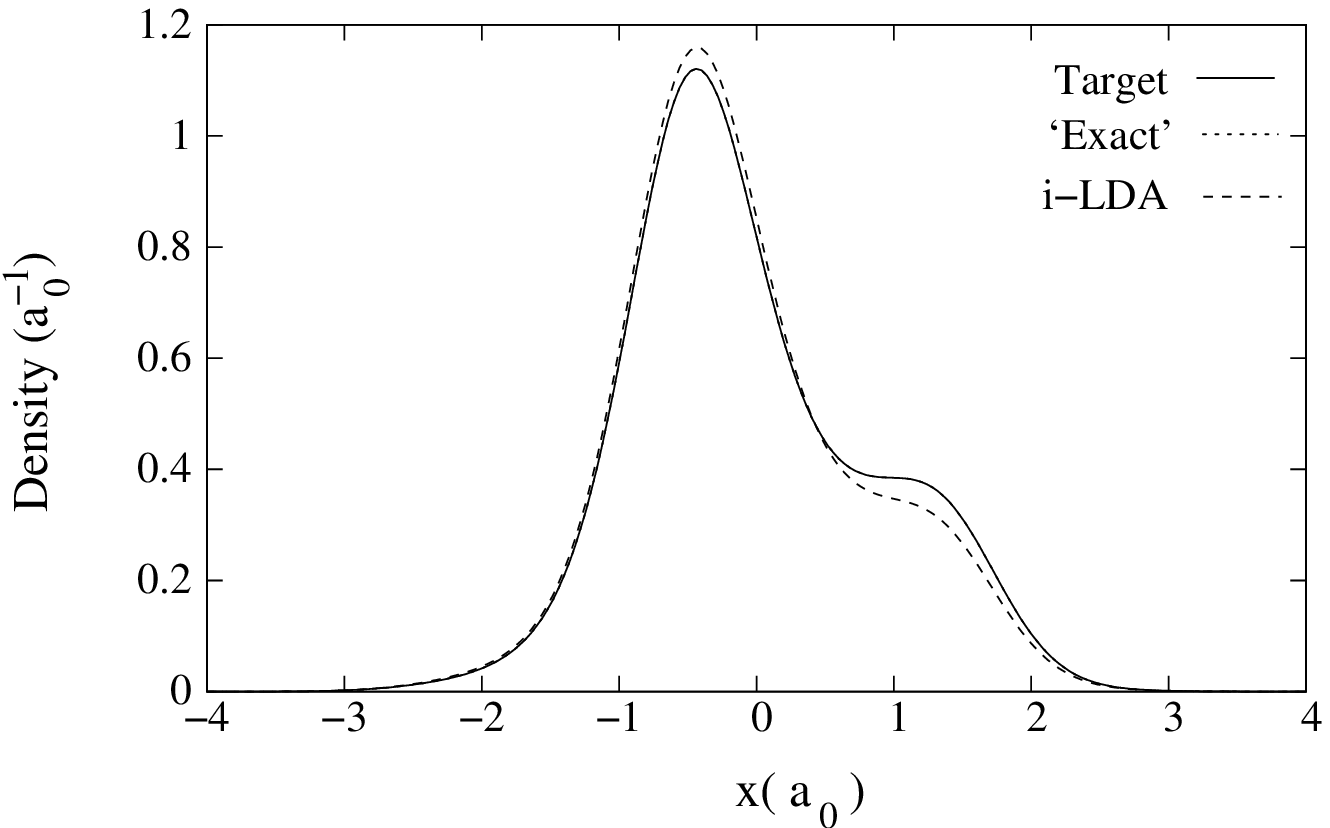}
\caption{The target density, `exact' density calculated using the `exact' potential arising from the iterative scheme and the LDA density obtained using the target potential}\label{fig:LDAsintypedensity}
\end{minipage}\hspace{2pc}%
\begin{minipage}{18pc}
\includegraphics[width=18pc]{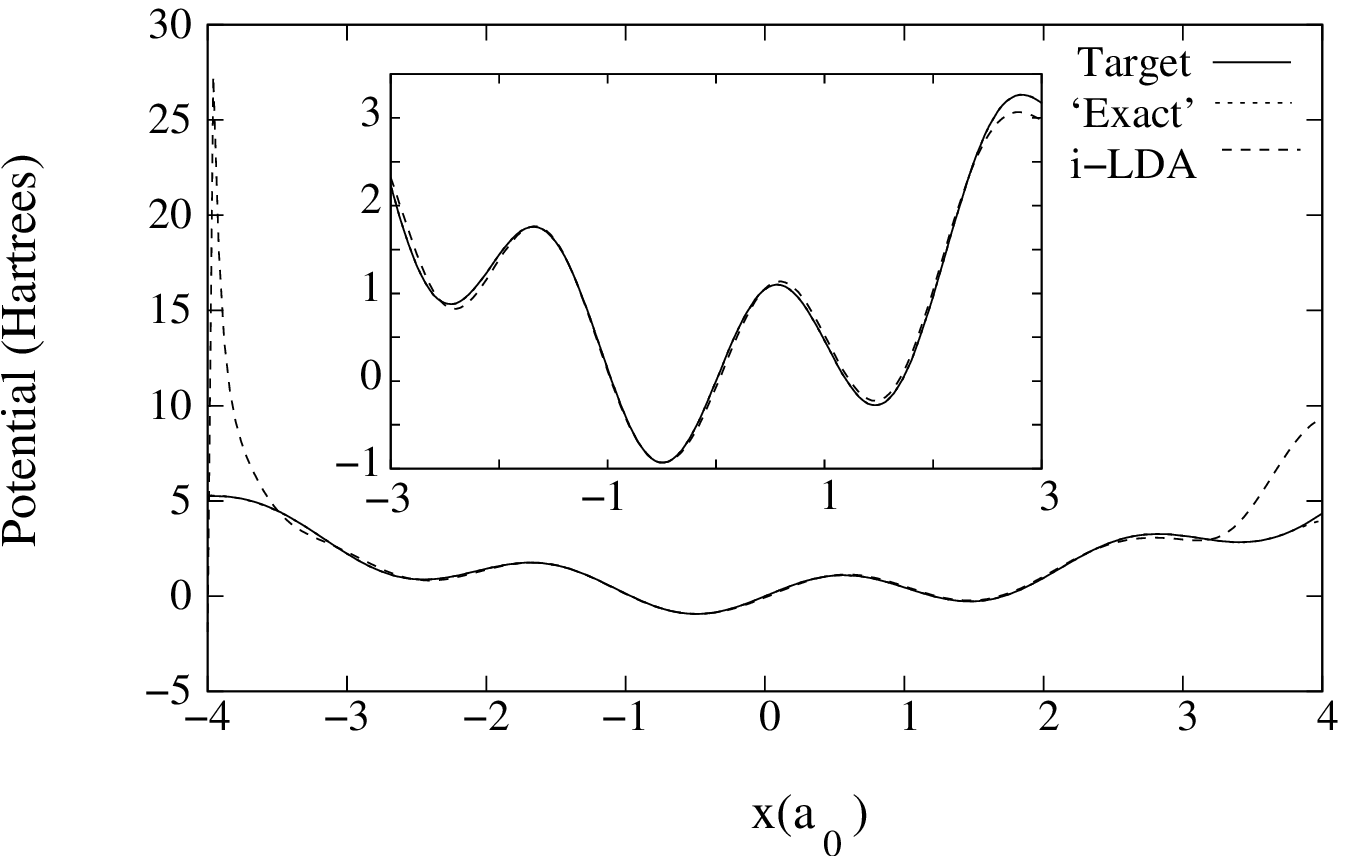}
 \caption{The target potential and the `exact' and i-LDA potential resulting from the iterative scheme.\\}\label{fig:iLDAsintypepot}
\end{minipage} 
\end{figure}

 We see in Fig.~\ref{fig:LDAsintypedensity} that the density has been reproduced almost exactly ($\sim 8.5\times 10^{-6}$ error) by using Eq.~\ref{eq:bothscheme} with $n^{target}$.  
We shift the resulting `exact' potential by a constant  so it gives the same energy as $v^{target}_{ext}$ and find that the `exact' potential is almost identical to the known potential: they cannot be distinguished by eye (Fig.~\ref{fig:iLDAsintypepot}) and differ only at the extremities where the density is low.  We then use $v^{target}_{ext}$  to calculate the LDA density which is a relative good match to the exact density, see  Fig.~\ref{fig:LDAsintypedensity}.  We now calculate the i-LDA potential which gives the LDA density to a high accuracy ($\sim 8.6\times 10^{-6}$ error). Fig.~\ref{fig:iLDAsintypepot} shows that the exact and i-LDA potentials are very different at the extremities but this is relatively unimportant as the density is negligible in this region.  An enlarged view of the region where 
the density is non-negligible (inset) shows that the i-LDA is a good match to the exact potential for most of the interval.

\section{System 2}
We now apply the iterative scheme to a density for which we do not know the potential.  We choose a flat-topped density which dies off as a Gaussian for values close to the box boundary as the target density.

\begin{figure}[!ht]
\begin{minipage}{18pc}
\includegraphics[width=18pc]{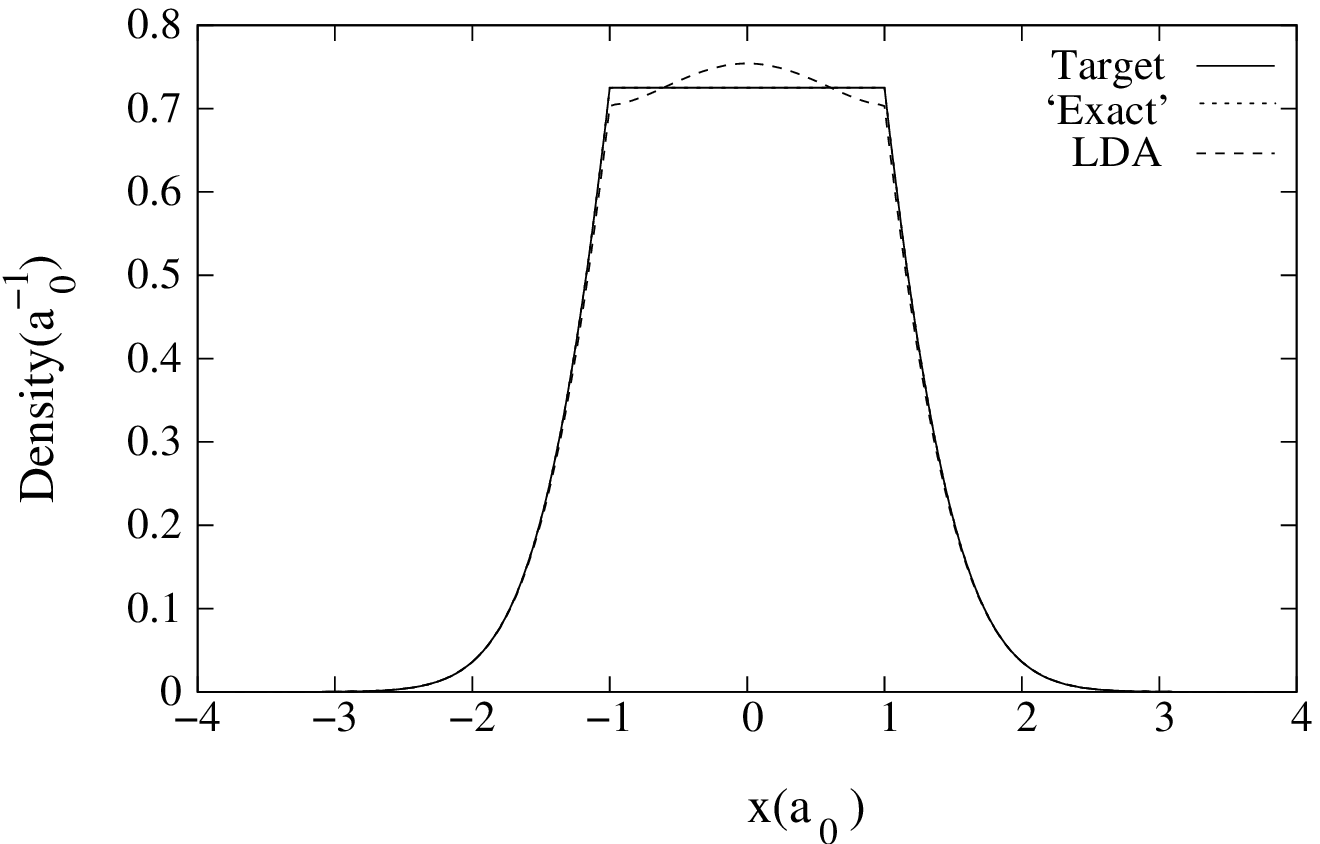}
\caption{The target density, `exact' density using the `exact' potential arising from the iterative scheme and the LDA density using this potential}\label{fig:flattopdensities}
\end{minipage}\hspace{2pc}%
\begin{minipage}{18pc}
\includegraphics[width=18pc]{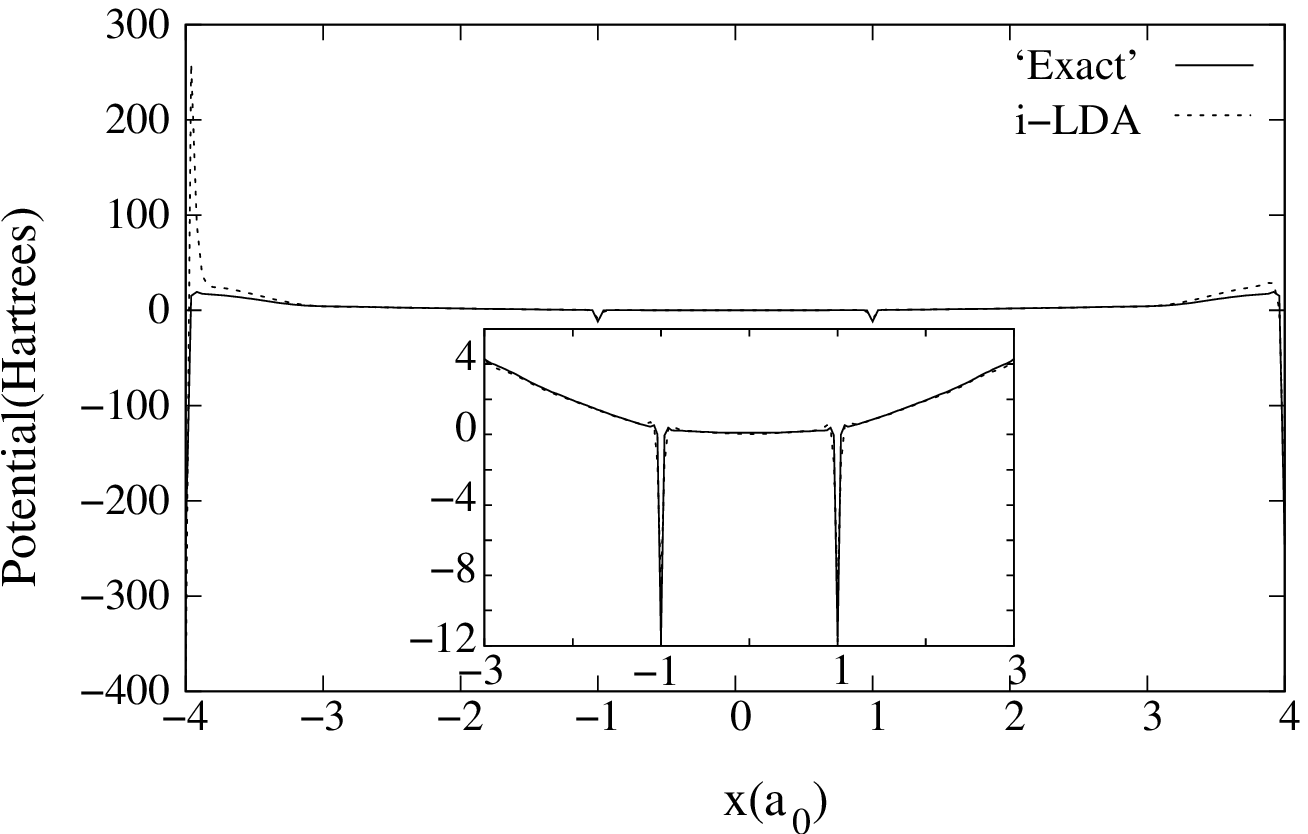}
 \caption{The `exact' and i-LDA potential for a density with a flat top.\\
\\}\label{fig:iLDAflattoppot}
\end{minipage} 
\end{figure}

We see in Fig.~\ref{fig:flattopdensities} that the target density is reproduced well ($\sim 5.7\times 10^{-5}$ error) by the `exact' potential shown in Fig.~\ref{fig:iLDAflattoppot}.  This potential is then used to calculate the LDA density which is shown in Fig.~\ref{fig:flattopdensities} and is a good match except in the flat region.  The i-LDA potential gives the LDA density with $\sim 1.5\times10^{-4}$ error and it is very different from the `exact' potential at the range boundaries. Here though the density is almost zero so the accuracy of the potential would not be expected to be high in this region.  For areas where the density is non-negligible we see that the i-LDA potential is similar to the `exact'.

%We note that if we try to find an almost rectangular density we encounter Gibbs type phenomenon in the density  ($Error=5.7332\times 10^{-3}$) see Fig~\ref%{fig:almostsquare}

%\begin{figure}[ht]\centering
%  \includegraphics[width=.4\textwidth]{almostsquare.eps}
%  \caption{The final trial density resulting from the iterative scheme compared to the %exact density.}\label{fig:almostsquare}
%\end{figure}

\section{Entanglement}
We may use the i-LDA system to calculate the electronic spatial entanglement corresponding to the LDA\cite{COE}.  We quantify the spatial entanglement using the Von Neumann entropy of the reduced density matrix $S=-Tr \rho_{\text{red}} \log_{2}\rho_{\text{red}}$  which we calculate by expressing $\rho_{\text{red}}$ in terms of the basis and diagonalising. Results are shown in Table \ref{tbl:energies}.
They show that the 1D LDA energies are very accurate  for the densities considered here, corroborating the results in \cite{MAGYAR04}. Also `exact' and LDA spatial entanglement are fairly close.  This is interesting as in earlier work\cite{COE} approximations to the LDA entanglement for the $3D$ Hooke's atom suggested a much less similar entanglement especially for values of S as large as found here.  We note though that for a confining potential with strength $\omega=0.5$ the $3D$ system was deemed as low interacting, with entanglement $S=0.0228$, and could be modelled well by the LDA.  However for  $1D$  Hooke's atom and $\omega=0.5$ we find $S=0.194$ -- similar to the magnitude of the entanglement found in this paper -- suggesting that the densities we investigated here also correspond to not too strongly interacting systems. This implies that LDA should be accurate for the corresponding entanglement, which supports our findings.
\begin{table}
\begin{center}
\begin{tabular}{|| c | c | c | c | c ||}
\hline
  \: &   Exact Energy  & LDA Energy &   Exact Entanglement & LDA  Entanglement  \\ 
\hline
System $1$ & $0.832$ & $0.821$ & $0.161$ & $0.140$ \\
\hline
System $2$ & $0.683$  &  $0.692$ & $0.206$  &  $0.194$ \\
\hline
\end{tabular}
\end{center}
\caption{Table showing the energy of the exact and LDA system in Hartrees and the exact and LDA spatial entanglement quantified by $S$ the Von Neumann entropy of the reduced density matrix.}\label{tbl:energies}
\end{table}

We have demonstrated that the scheme Eq.~\ref{eq:bothscheme} can be used to find a potential which reproduces a target ground state density of an interacting system.  In addition the scheme may be used to find the entanglement corresponding to the LDA.
\section*{Acknowledgements} We thank K. Capelle and E. R\"as\"anen for fruitful discussions and acknowledge funding from EPSRC grant EP/F016719/1.

\end{document}